# Excitation enhancement of a quantum dot coupled to a plasmonic antenna

*Esteban Bermúdez Ureña*, *Mark P. Kreuzer*, *Stella Itzhakov*, *Hervé Rigneault*, *Romain Quidant*, *Dan Oron*, and *Jérôme Wenger\**

[*]     Dr J. Wenger Corresponding-Author,
Institut Fresnel, Aix-Marseille Université, CNRS, Ecole Centrale Marseille,
Campus de St Jérôme, 13397 Marseille, France
E-mail: jerome.wenger@fresnel.fr
        E. Bermúdez Ureña, Dr M. P. Kreuzer, Prof. R. Quidant,
ICFO-Institut de Ciences Fotoniques,
Mediterranean Technology Park, 08860 Castelldefels, Spain
        S. Itzhakov, Dr D. Oron,
Department of Physics of Complex Systems, Weizmann Institute of Science,
Rehovot 76100, Israel
        Dr H. Rigneault,
Institut Fresnel, Aix-Marseille Université, CNRS, Ecole Centrale Marseille,
Campus de St Jérôme, 13397 Marseille, France
        Prof. R. Quidant,
Institucio Catalana de Recerca i Estudis Avançats,
Barcelona, 08010, Spain

Plasmonic antennas excited at resonance create highly enhanced local fields,[1] which are key for surface-enhanced Raman scattering,[2,3] biosensing,[4,5] and nonlinear photoemission down to nanoscale volumes.[6,7] Plasmonic antennas also enable the control of the optical emission from single quantum emitters.[8-18] However, a major challenge in coupling a single emitter to a plasmonic antenna is that the proximity of the quantum emitter to the metal results in energy transfer to the density fluctuations of the free electron gas that simultaneously enhances radiative emission and generates ohmic losses.[8-10] Depending on the experiment configuration and the balance between radiative and non-radiative decay rates, either luminescence enhancement or quenching is reported.[16-18] This apparent confusion arises as the luminescence signal combines modifications on excitation intensity, quantum yield, and collection efficiency. Moreover, there are only few reports experimentally quantifying the antenna's influence on excitation and emission.[11,14] New experimental methods are required to fully investigate the antenna response, and quantify separately the enhancement factors for excitation and emission.

Colloidal quantum dots (QDs) offer much wider possibilities than organic fluorophores for the investigation of the antenna-emitter coupling. The broad absorption spectrum of QDs can be used to probe the photoluminescence intensity on and off the antenna resonance to derive the excitation intensity enhancement.[19,20] Another approach takes advantage of sequential resonant photon absorption and multiply excited states occurring in QDs.[21-23] We use this property here to quantify the local excitation intensity enhancement. The transient photoluminescence dynamics of the QDs contain at least two contributions, respectively from the singly and doubly (and higher) excited states. An important feature is that the ratio of doubly to singly excited state photoluminescence amplitudes quantifies the local excitation intensity independently on the emission process.[24] Another advantage of this approach is that in the same measurement the single exciton lifetime and intensity can be measured, from which the emission gain can also be deduced, providing the full information about the antenna's influence.

Here we investigate the luminescence of a quantum dot deterministically coupled to a gold nanoantenna, and quantify the antenna's influence on the excitation intensity and the luminescence quantum yield separately. We report on monomer and dimer disk antennas with resonances close to the excitation and emission wavelengths. These antenna designs yield higher excitation enhancement but also higher quenching losses as compared to previous studies on polystyrene microspheres and subwavelength apertures.[24] Thus the separate investigation of excitation and emission processes is key to fully understand the antenna's influence on the QD luminescence.

Typical fabricated antennas are shown in **Figure 1**a. Each gold particle has a 90 nm diameter, 40 nm thickness, and the gap size is either 14 or 30 nm for the results reported here. The dimer antenna design requires the precise near-field coupling of the emitter to the gap antenna. To achieve this, we perform a two-step electron beam lithography process combined with chemical functionalization and binding of the Au structures and the QDs: the first lithography step defines the antenna structures on an ITO substrate, the second lithography step defines the area for chemical binding of the QDs (Figure 1b). [13] As a last step, custom-made core/shell/shell CdSe/CdS/ZnS QDs are immobilized on the functionalized areas, and the excess QDs are removed by a lift-off step of the remaining PMMA. The QDs have quasi-spherical shapes with 10 nm diameter (Figure 1c) and peak emission at 660 nm with 30 nm FWHM. Details about the antenna fabrication, QD synthesis, functionalization and chemical coupling route to the nanostructures are given in the Experimental Section and in the Supporting Information. Extinction spectra representative of the studied antennas are shown in Figure 1d (prior to the QD binding procedure, which red-shifts the resonances by about 20 nm). This ensures the nanostructures resonances are close to both the excitation wavelength and the spectral range for the luminescence detection. Given the broad spectral response of the antennas as compared to the 30 nm FWHM emission from the QD, the antenna is assumed to induce only marginal modifications to the luminescence spectrum.

**Figure 2**a displays confocal photoluminescence images corresponding to the case of a dimer antenna with excitation polarization parallel or perpendicular to the dimer long axis. Over 80% of antennas show QD photoluminescence, and about 50% show polarization sensitive emission (red circles on Figure 2a represent the antennas selected for further investigations). For these antennas, the photoluminescence ratio between parallel and perpendicular excitation is higher than 10. The QD luminescence also turns into a clear linear polarization parallel to the long axis of the antenna, with a degree of linear polarization of 0.8 indicating QD coupling to the antenna.[13]

Figure 2b presents typical photoluminescence time traces. For QDs lying on bare ITO substrate, the blinking dynamics are strongly suppressed as compared to QD on glass. We also observe a reduced exciton lifetime of 2.0 ns for QD on ITO while the exciton lifetime is 9.5 ns on glass. These features relate to energy transfer from the QD to the conductive ITO layer, as previously reported.[25-26] Interestingly, when the QD is coupled to a plasmonic antenna the blinking dynamics are partly retrieved while the luminescence lifetime is further reduced.[27] This is indicative of an enhancement of the QD emission rate upon coupling to the metal antenna.

Another signature of the near-field coupling between dimer antenna and QD is found in the radiation pattern. The images in Figure 2c record the photoluminescence intensity distribution on the back focal plane (Fourier plane or momentum space) of the high numerical aperture (NA) objective, and contain the directions of emission toward the substrate. Two distinct circles are seen in the polar angle. The outer circle is the maximum collection angle of the 1.2NA water-immersion objective (64°). The inner circle is the critical angle for the glass-air interface (NA=1 or 41.1°), where a dipole close to a glass interface is expected to emit with a sharp maximum. The radiation patterns of a QD on ITO (left column) and of a QD coupled to a single gold particle (center) are isotropic in the azimuthal angle, as a direct consequence of the QD degenerate transition dipole moment.[28,29] When the QD is coupled to a dimer gap antenna (Figure 2c, right column), the radiation pattern changes dramatically and transforms to that of a linear dipole horizontally aligned respective to the interface. Hence the antenna mode fully determines the QD radiation pattern, which is a further evidence of coupling between the QD and the antenna.[13,30,31]

As a consequence of the strong quantum confinement of the free charge carriers, QDs can undergo sequential resonant photon absorption, and sustain multiply excited states.[21] Below photoluminescence saturation, the singly excited (X) state mostly results from the absorption of a single photon while the doubly excited state (BX) state is created after the sequential absorption of two photons (**Figure 3**a). Immediately after pulsed photoexcitation, the X and BX relative populations thus scale linearly and quadratically (respectively) with the excitation intensity. The X and BX populations can be distinguished by monitoring the QD's transient emission dynamics, [22] since the X state radiative lifetime is on the order of a few nanoseconds while the BX lifetime is dominated by Auger recombination and ranges between ten and a few hundreds of picoseconds (Figure 3a). The time correlated single photon counting trace $s(t)$ is decomposed as a sum of two exponentials:

$$s(t) = a_X \exp(-t/\tau_X) + a_{BX} \exp(-t/\tau_{BX}) \qquad (1)$$

where $a_X$ and $a_{BX}$ are the amplitudes of characteristic decay times $\tau_X$ and $\tau_{BX}$, respectively.

The amplitude $a_X$ relates to the singly excited X state, while $a_{BX}$ relates to the doubly excited BX state, as demonstrated later by the excitation power dependence. Contrarily to earlier work,[24] the X state decay of the QDs used in this study is well modeled by a single exponential (see Supporting Information). The $a_X$ and $a_{BX}$ amplitudes can be expressed as function of the radiative decay rate $k_i^{rad}$ for the $i$-th mulitexciton state and the average number $N_{abs}$ of absorbed excitation photons per QD per pulse: [22,24]

$$a_X \propto k_X^{rad} N_{abs} \qquad (2)$$

$$a_{BX} \propto k_X^{rad} (r-1) N_{abs}^2 \qquad (3)$$

Here $r = k_{BX}^{rad} / k_X^{rad}$ is the degeneracy factor of the BX state, which typically amounts to a constant value $r \sim 2\text{-}3$ fixed by the nature of the QD.[22,32]

Equations (2) and (3) contain two important points. First, both $a_X$ and $a_{BX}$ are linearly proportional to the radiative rate $k_X^{rad}$. Hence both $a_X$ and $a_{BX}$ sense the same emission rate enhancement on plasmonic antennas. Second, $a_X$ bears a linear dependence with $N_{abs}$ and the local excitation intensity while $a_{BX}$ has a quadratic dependence. While computing the ratio $a_{BX}/a_X$ the emission rate contribution cancels out, and only a term proportional to the local excitation intensity remains. As *r* is a constant, $a_{BX}/a_X$ is thus a direct probe of the local excitation intensity. An increase in this ratio on a nanoantenna as compared to a flat interface is a direct demonstration of an increased local excitation intensity, independently on the number of emitters involved and the emission enhancement.

Figure 3b-d presents experimental decay traces and the corresponding bi-exponential fit according to Equation (1). These decay traces are normalized to better reveal the modifications of the decay dynamics and the increased relative weight of the fast BX transient component as the excitation power is raised. For the QD on ITO, the exciton lifetime amounts to $\tau_X$ =2.0 ns, while for the QD coupled to a monomer antenna and for the QD coupled to a dimer antenna we find $\tau_X$ =0.18 ns after deconvolution from the instrument response function.[14] This modified exciton lifetime corresponds to a lifetime reduction or excitonic decay rate enhancement of 11.1 when the QD is coupled to the antenna. For all samples tested, the biexciton lifetime is found to $\tau_{BX}$ =0.12 ns which is presently limited by the instrument response function. This short lifetime confirms that the Auger recombination route strongly affects the BX to X decay. The excitonic and biexcitonic amplitudes $a_X$ and $a_{BX}$ are displayed in the last column of Figure 3b-d. The excitonic amplitude $a_X$ is found to grow linearly with the excitation intensity, while the biexcitonic amplitude $a_{BX}$ grows quadratically, as expected from Equations (2) and (3). The evolution of $a_{BX}$ with the excitation power confirms that the fast transient component in the decay trace corresponds to the bi-excitonic process BX.

From the data set in Figure 3b-d, we compute the ratio $a_{BX}/a_X$, which follows a linear trend as the excitation power is increased (**Figure 4**a). For all excitation powers, the $a_{BX}/a_X$ ratio on plasmonic antennas exceeds the one found for the ITO reference. As demonstrated by Equations (2) and (3), this effect is directly related to the excitation intensity amplification induced by the antenna. To quantify the excitation intensity enhancement, we take the ratio of the slopes in Figure 4a, which provides a better estimate with a typical relative uncertainty of 10% for the excitation enhancement factors on the different selected antennas.

The orange bars in Figure 4b summarize our main results. The excitation enhancement factor starts at 5.1 for a single disk antenna. For the dimer antennas, as the gap size is reduced, the electromagnetic coupling between the gold particles is increased.[11,31] Our observations confirm this trend by an increase in the excitation enhancement from 13.1 to 15.9 when the gap is reduced from 30 to 14 nm. Let us stress that these enhancement factors for the excitation intensity sensed by the QD are independent on the number of emitters involved in the luminescence signal and on the emission properties. The experimental results also stand in good correspondence with numerical simulations based on the finite-difference time-domain FDTD method (see the Experimental Section for details). The only exception concerns the case with highest enhancement (dimer with 14 nm gap, parallel excitation), for which nanofabrication deficiencies, non-ideal QD positioning and more complex photodynamics have a non-negligible influence.

The QDs used in this work are custom synthetized to bear a high BX population. If this is an advantage to determine the $a_{BX}/a_X$ ratio, it also has the negative consequence to blur any photon antibunching experiment because of the emission of two photons within the BX state decay. Hence standard Hanbury-Brown-Twiss experiments cannot be successfully implemented to guarantee that there is a single QD under investigation. Instead of this, clues for the single emitter on the nanoantenna are brought by (i) the blinking dynamics down to the background level (see Figure 2b for typical time traces), and (ii) the luminescence intensity level on the selected antennas as compared to other antennas where most likely more than one QD is present (compare the intensity levels in Figure 2a). For the reference on the ITO substrate, we focus on an area with very low QD coverage density, where each bright spot is diffraction-limited and the intensity of bright spots follow a Poisson-like distribution. Selecting the spots with minimum intensities promotes the cases where most likely a single QD is present. Hence to quantify the luminescence enhancement, we compute the ratio of the average levels of luminescence from the time traces, and assume the detected signal stems from a single QD.

Figure 4c summarizes the enhancement factors found for the photoluminescence intensity. While a value of 2.7x is found for the single disk antenna, the luminescence is only enhanced by a factor of 1.3x for a dimer gap antenna with 14 nm gap and parallel excitation, although this configuration leads to a 15.9 excitation intensity enhancement. The situation is even worse for the dimer with perpendicular polarization orientation, where the luminescence factors drops to 0.25x the ITO reference. We remind that for the different cases the exciton decay rate is increased by a factor 11.1x as compared to the ITO reference. The fact that the luminescence enhancement is smaller than the decay rate enhancement and the excitation intensity enhancement is an indication of quenching, i.e. the non-radiative energy decays take the lead over the radiative routes.

In the excitation regime below photoluminescence saturation, the luminescence enhancement is proportional to the gains in collection efficiency, quantum yield and excitation intensity. Thus from the measurements of the excitation and luminescence enhancement factors (Figure 4b,c), we compute the enhancement factor for the QD quantum yield multiplied by the collection efficiency. The results are displayed in Figure 4d. For all configurations, values below 1 are obtained, which further evidence luminescence quenching. Our observations support the fact that the quenching losses also increase as the gap size is reduced:[18] antennas with gaps of 14 or 30 nm are found to provide almost the same quantum yield enhancement, although the excitation enhancement was found to be significantly higher in the case of the 14 nm gap. A remarkable feature of our study is that excitation intensity enhancement can be recorded despite this quenching phenomenon, and that near-field intensity information can be extracted from emission even in the presence of strong non-radiative losses.

To conclude, this work provides new routes to experimentally investigate the physics of optical antennas, and optimize the excitation and emission processes independently for the future development of bright single-photon sources and biochemical sensors. Positioning a quantum dot respective to an optical antenna provides also an interesting approach to investigate the exciton-photon interaction beyond the classical dipole approximation.[33,34]

*Experimental*

*Synthesis of CdSe/CdS/ZnS QDs:* 1-octadecene (ODE; tech. 90%), trioctylphosphine (TOP; 90%), cadmium oxide (CdO; 99.99+%), sulfur powder (99.98%), selenium powder (100 mesh, 99.5+%), zinc acetate (Zn(Ac)2; 99.99%), oleic acid (OA, tech. 90%), and hexadecylamine (HDA, tech. 90%) were ordered from Aldrich. Toluene, acetone, chloroform and methanol were ordered from Bio Lab. The synthesis of CdSe/CdS/ZnS nanocrystals was performed by following the procedure suggested by Peng and coworkers with minor modifications [35]. The whole synthesis was carried out using a Schlenk technique. A mixture of CdO (13 mg, 0.1 mmol), OA (0.3 ml), HDA (1.3 ml) and ODE (4 ml) was dried and degassed under vacuum at 120°C for 10 min. in a 50 ml three-neck flask. Following, the solution was heated under argon to 280°C, and the stock solution of TOP:Se prepared by dissolving elemental selenium (8 mg, 0.1 mmol) in a 2 ml TOP was quickly injected to the hot solution under vigorous stirring. The growth temperature was then reduced to 250°C until no significant growth of dots was detected (about 1 hr.). At this stage the diameter of CdSe cores is 4.6 nm with the first absorption peak at 600 nm. Additional one layer of CdSe, as well as CdS and ZnS shells were grown using a layer-by-layer growth technique in a one pot synthesis. The injection solutions used for the CdSe layer are 0.1 M Se in TOP (dissolved at room temperature) and 0.1 M Cd oleate in ODE (the OA to CdO molar ratio is 8:1, and ODE is added to reach the final concentration of 0.1 M). CdO, OA, and ODE were degassed under vacuum at 100°C for 30 min. Following, the temperature was increased under argon to 260°C to get a clear solution of 0.1 M Cd oleate in ODE. This precursor solution was cooled and used at room temperature. The CdS precursor solutions are 0.1 M cadmium oleate and 0.1 M sulfur in ODE. The injection solutions used for the ZnS shell growth are 0.1 M zinc oleate in ODE and 0.1 M sulfur in ODE. Zinc oleate is prepared similarly to the cadmium oleate, but instead of CdO, Zn acetate was used, and the solution was left at 275°C for 1.5 hr. to become clear. For each injection, a calculated amount of a given injection solution was taken and injected in a dropwise manner to the solution containing CdSe cores. One layer of CdSe was grown at 300°C, four layers of CdS were grown at 280-300°C, two layers of ZnS, and one additional layer of Zn were grown at 280°C. After each shell growth, the nanocrystals were annealed at 300°C for 20 min. Following the last annealing, the solution was cooled to room temperature and stored in a freezer till precipitation procedure and ligand exchange.

*Precipitation procedure:* Separation of unreacted precursors from the nanocrystals was performed by extraction followed by precipitation (twice for each process). For extraction, a bit of toluene and a large quantity of methanol were added until phase separation. After centrifugation, the upper phase was discarded, and the process was repeated. To precipitate nanocrystals after the second extraction, acetone was added to get a turbid dispersion. The precipitate was re-dispersed in toluene, and the precipitation process was repeated one more time.

*Ligand Exchange of Quantum Dots:* QDs as synthesized were redispersed in chloroform and adjusted to have an optical density ca. 1 at the first exciton peak (617 nm). 5 mg of alpha-carboxy-omega-mercapto poly(ethylene glycol)hydrochloride (Iris-Biotech) with a molecular weight of 3163 were dissolved in a 50 mg/ml solution of potassium hydroxide in water, pH 10 and added to the organic QDs. The vial was shaken and allowed the phases to separate. At this point, methanol was added to the mixture and again shaken briefly. The modified QDs had now passed into the methanol/water layer. Chloroform was removed and the aqueous layer again washed with fresh chloroform. Sodium chloride was now added to the aqueous phase to remove any unreacted carboxy-PEG followed by 2x excess of methanol. This mixture was centrifuged at 6000 rpm for 15min and allowed the formation of a pellet. The supernatant was decanted and the pellet redispersed in water, pH 5-6 with gentle sonication for a few seconds.

*Chemical Functionalization of gold nanoantennas:* Gold nanoantennas were defined by electron beam lithography in a FEI InspectF50 system at 30 keV acceleration voltage (see Supporting Information for details). The nanoantenna sample (substrate and PMMA) that had previously been exposed to e-beam lithography and resist development in order to define hole areas was immersed in a 10 mM solution of mercaptoundecanoic acid in ethanol and left overnight at room temperature (RT). After 18 hours, the sample was removed, rinsed in ethanol and then water before drying in a flow of nitrogen ($N_2$). Solutions of 10 mM EDC (N-(3-dimethylaminopropyl)-N'-ethylcarbodiimide hydrochloride) and NHS (N-hydroxysuccinimide) were prepared in 20 mM MES buffer, pH 5.5 and a 1:1 (v/v) mixture was added to the PMMA coated substrate and allowed to react for 25 min at RT. Thereafter the sample was rinsed in MES buffer before applying a 1% solution of bovine serum albumin (BSA) in PBS buffer, pH 7.5 for 1 hr at RT. Prior to completion, the modified QDs with the acid group were also activated in a similar fashion. Briefly, 20 µl of QDs were added to 20 µl of MES. Then 3.2 µl of EDC and 2 µl of NHS (both 100 mg/ml in MES, pH 5.5) were added and allowed to interact for 25 min at RT. In the meantime the substrate was rinsed from excess of BSA and dried in $N_2$. The activated QDs were now diluted to 300 µl in PBS, pH 7.5 and all solution was dropped onto the dry substrate and left overnight at RT in the dark. The following day the drop was removed and rinsed before performing standard lift-off in acetone to remove PMMA and any excess QDs not chemically attached to the BSA protein.

*Confocal photoluminescence measurements:* Photoluminescence experiments are performed on a confocal microscope with a 40x, NA 1.2 water-immersion objective. The excitation source is a pulsed laser diode operating at 636 nm with 50 ps pulse duration and 40 MHz repetition rate. The laser diode profile is spatially filtered by a single mode fiber to provide diffraction-limited focusing (waist calibrated to 275 nm). QD luminescence is collected using the same microscope objective, and detected by fast avalanche photodiodes after passing a 30 µm confocal pinhole. The luminescence intensity is integrated over the 650-690 nm range of the QD emission. The detected count rates do not exceed 1% of the repetition rate, avoiding photon pile-up artifacts. Lastly, transient emission dynamics are analyzed by a fast time-correlated single photon counting module (PicoQuant PicoHarp 300). The overall temporal resolution is 120 ps.

*Extinction spectroscopy*: The studied sample also comprises denser arrays of the same antennas to measure the extinction resonance of each structure. The optical setup consists of a standard microscope in bright field configuration. The illumination is performed from the bottom side of the sample by a 100 W halogen lamp with a linear polarizer and a bright field condenser (0.1NA). The transmitted light is collected with a bright field objective (10x, 0.25 NA) and passed through a beam splitter into a CCD camera for alignment and to a spectrometer (Andor, Shamrock) via an optical fiber (200 µm diameter).

*Numerical simulations*: Three-dimensional numerical modelling on nanoantennas is based on the finite-difference time-domain FDTD method using Rsoft Fullwave version 6.0. The model considers a computational space of 0.3 x 0.3 x 0.2 $\mu m^3$ with 0.7 nm mesh size and perfect matched layers boundary conditions on all faces. The gold antenna (refractive index 0.183+2.974i) is on a glass substrate (refractive index 1.52), the upper medium is air. The QD is modeled by a 10 nm sphere with refractive index 2.47, which corresponds to the average refractive indexes of the QD components at the excitation wavelength. For the dimer antennas, the QD is positioned in the exact center of the gap, while for the monomer antenna, the QD center is set at a distance of 8 nm from the disk surface. Excitation at 636 nm is launched incoming from the glass side. To quantify the excitation enhancement, we average the intensity at the surface of the QD sphere, and normalize this value by the average intensity found on the QD without the antenna.

*Acknowledgements*
The research leading to these results has received funding from the European Research Council under the European Union's Seventh Framework Programme (FP7/2007-2013) / ERC Grant agreements 258221 (SINSLIM), 259196 (Plasmolight), and 278242 (ExtendFRET), the Spanish Ministry of Sciences through grants FIS2010 12834 and CSD2007-046-Nano-Light.es, the Fundacio Privada CELLEX, the Israeli Ministry of science & Technology and the Provence-Alpes-Côte d'Azur Region. We thank Giorgio Volpe for valuable input and fruitful discussions. This research has been conducted in the scope of the CNRS-Weizmann NaBi European associated laboratory.

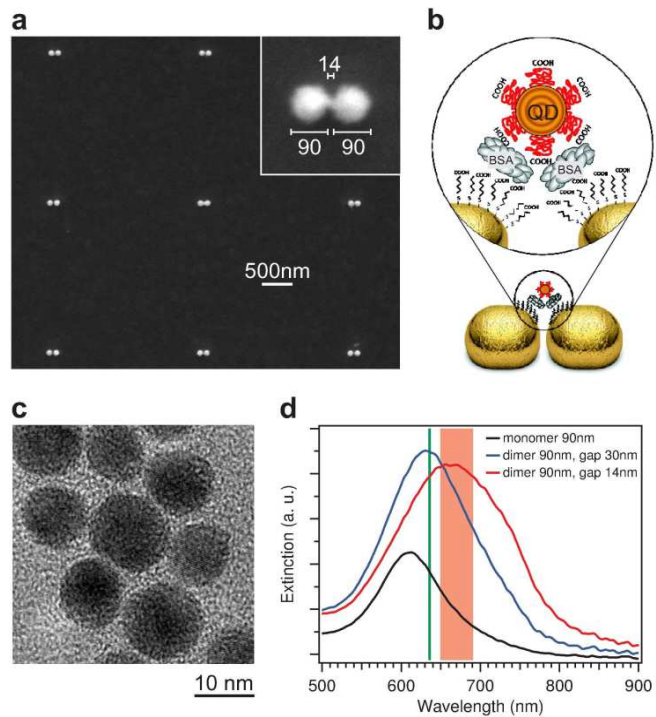

**Figure 1.** (a) Scanning electron microscopy image of fabricated dimer gap antennas after the QD deposition procedure. Indicated distances are in nm. A QD is deterministically attached in the gap region after double lithography process, as depicted in (b). (c) Transmission electron microscopy (TEM) image of the fabricated core/shell/shell CdSe/CdS/ZnS QDs. (d) Extinction spectra representative of the studied antennas. Monomers of 90 nm diameter (black), and dimers of 90 nm particles with gaps of 30 nm (blue) and 14 nm (red). The positions of the excitation wavelength and collection emission window are depicted with the green line and red frame respectively.

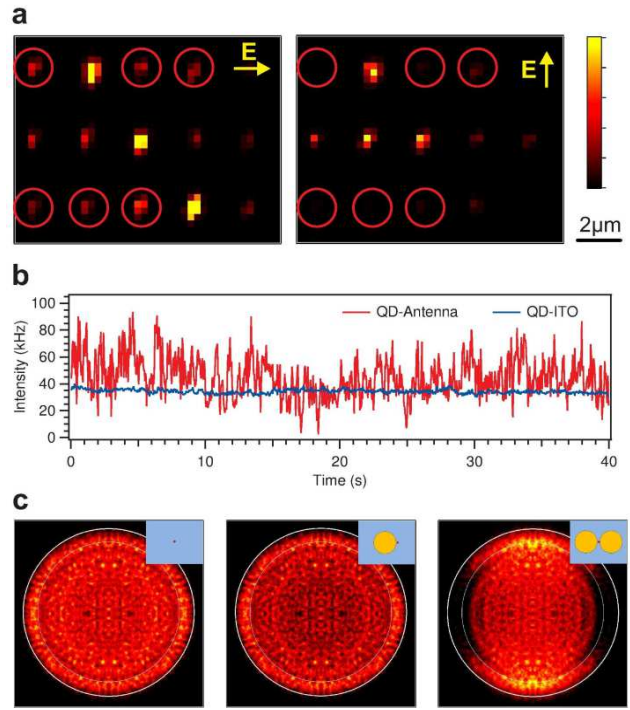

**Figure 2.** (a) Confocal photoluminescence images for excitation polarization parallel (left) and perpendicular (right) to the dimer long axis. The gap size is 14 nm here. Red circles indicate selected antennas for photoluminescence studies. (b) Photoluminescence time trace for a QD coupled to a dimer antenna (red line) and on ITO (blue line), average excitation power is 10 µW. (c) Radiation patterns (back focal plane image of the 1.2NA objective) from a QD on ITO substrate (left), a QD coupled to a single gold particle (center), and a QD coupled to a dimer gap antenna (right).

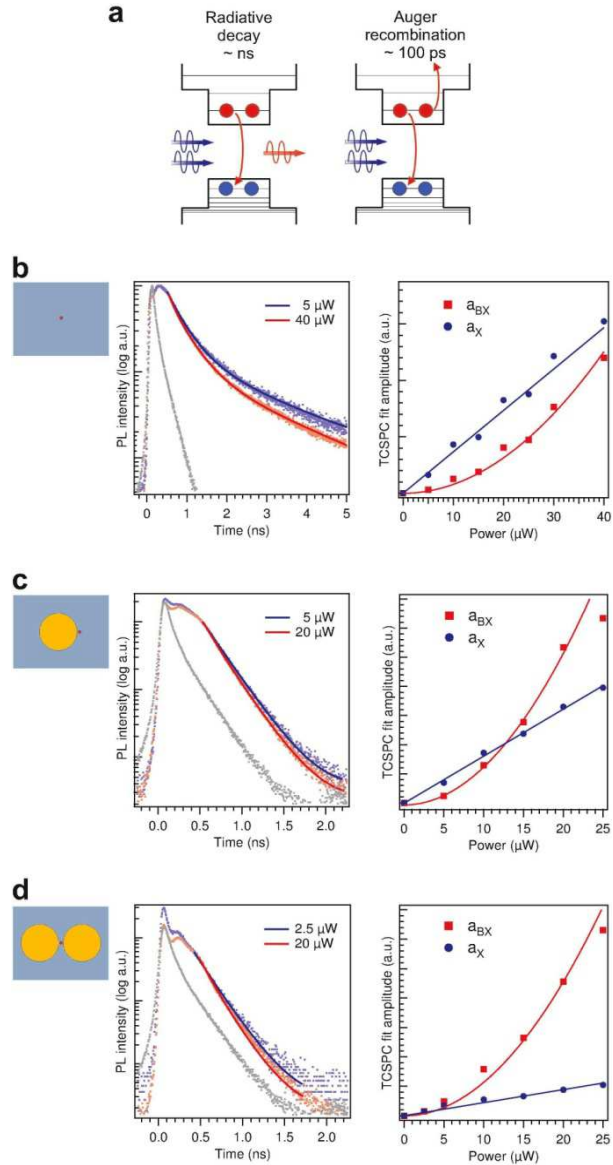

**Figure 3.** (a) Scheme of the decay routes for biexcitonic BX state. (b-d) Luminescence decay curves and bi-exponential fit amplitudes $a_X$ and $a_{BX}$ for the different cases of a QD on ITO substrate (b), a QD coupled to a single gold particle (c), and a QD coupled to a dimer antenna with 14 nm gap (d). The light gray decay trace shows the instrument response function. Note the different scale for the excitation power in (b) vs. (c-d).

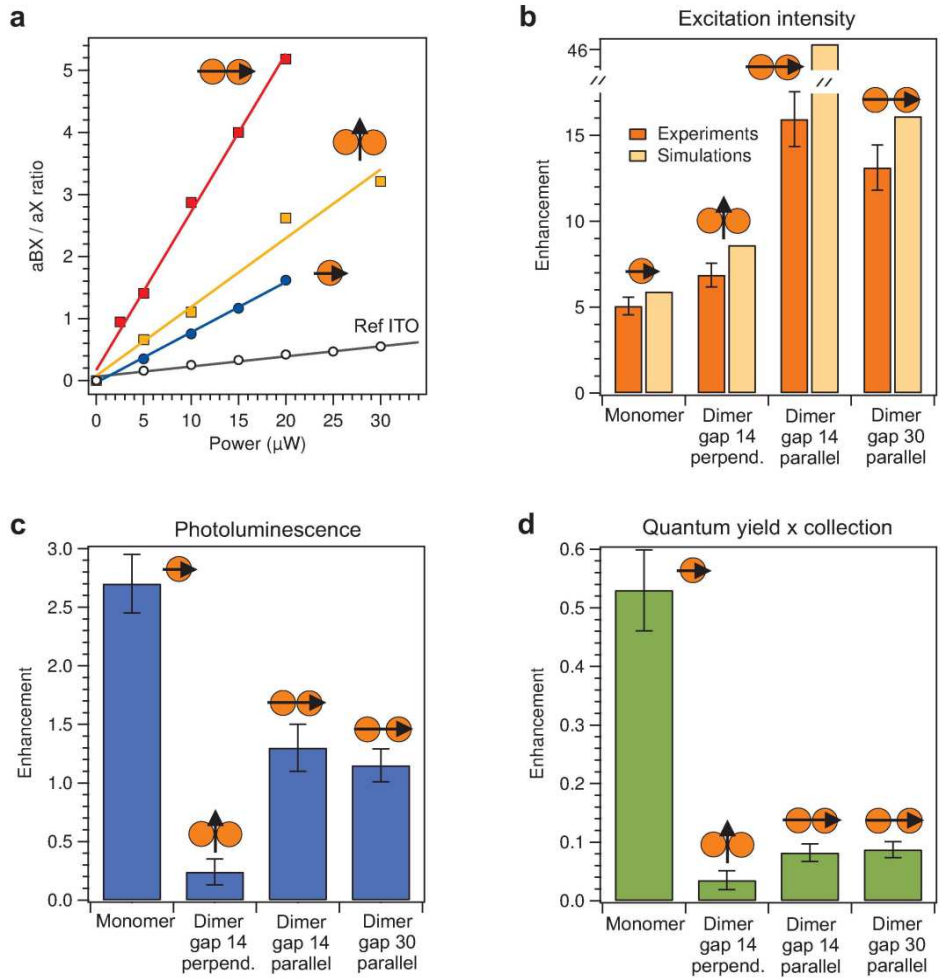

**Figure 4.** (a) Ratio of biexcitonic / excitonic amplitudes $a_{BX}/a_X$ versus excitation power for a QD on ITO (black), a QD coupled to a single gold particle (blue), and a QD coupled to a dimer gap antenna with 14 nm gap and parallel (red) or perpendicular (orange) excitation polarization. (b) Excitation enhancement deduced from the data slopes in (a). (c) Integrated photoluminescence enhancement. (d) Quantum yield and collection efficiency deduced from the data in (b) and (c).

# Supporting information for
# Excitation enhancement of a quantum dot coupled to a plasmonic antenna

*Esteban Bermúdez Ureña, Mark P. Kreuzer, Stella Itzhakov, Hervé Rigneault, Romain Quidant, Dan Oron, and Jérôme Wenger*

**TEM**

Five microliters of the sample were applied to a 400-mesh copper grid coated with nitrocellulose and carbon. Samples were blotted after ~10 sec. and dried in air. TEM was performed on a Philips T12 transmission electron microscope operated at 120 kV and equipped with a TVIPS CCD digital camera.

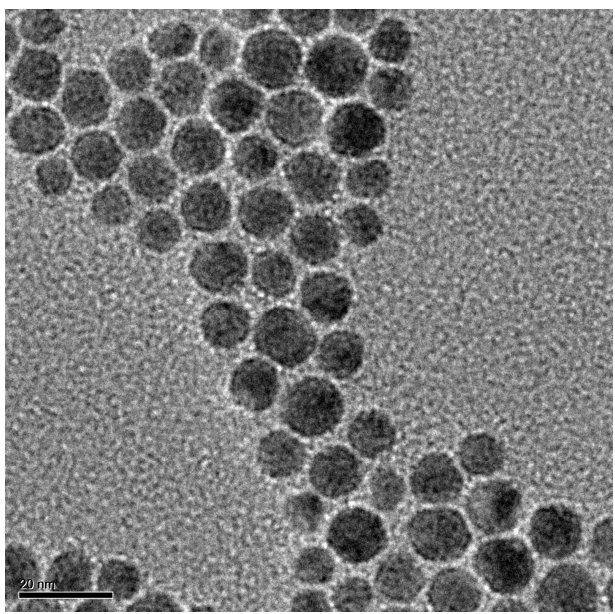

**Figure S1.** TEM image of spherical CdSe/CdS/ZnS quantum dots. The scale bar is 20 nm.

**Excitation and emission spectra**

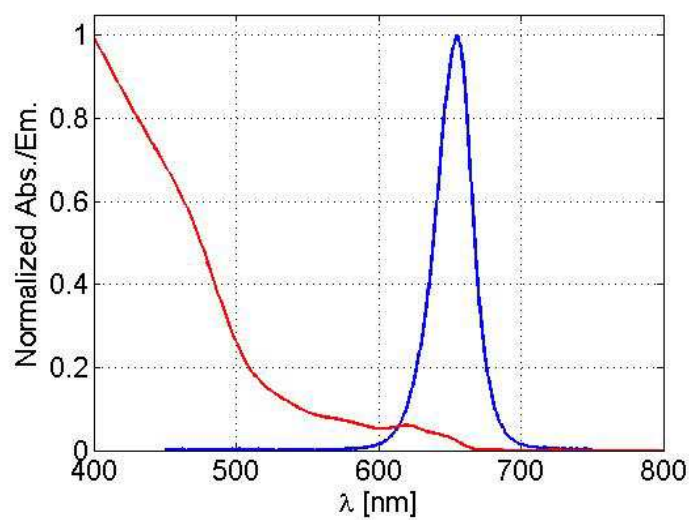

**Figure S2**. Absorption (red) and emission (blue) spectra of the spherical CdSe/CdS/ZnS quantum dots in solution.

**Ligand Exchange of Quantum Dots**

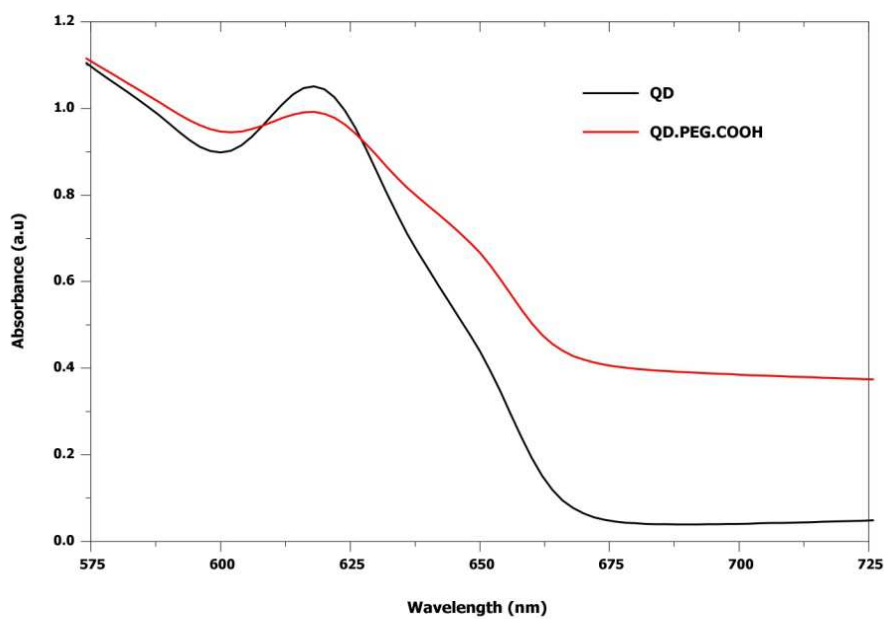

**Figure S3.** Absorbance spectra showing QDs as synthesized and once conjugated to carboxy-mercapto poly(ethylene glycol).

**Photoluminescence decay trace of quantum dots in solution**

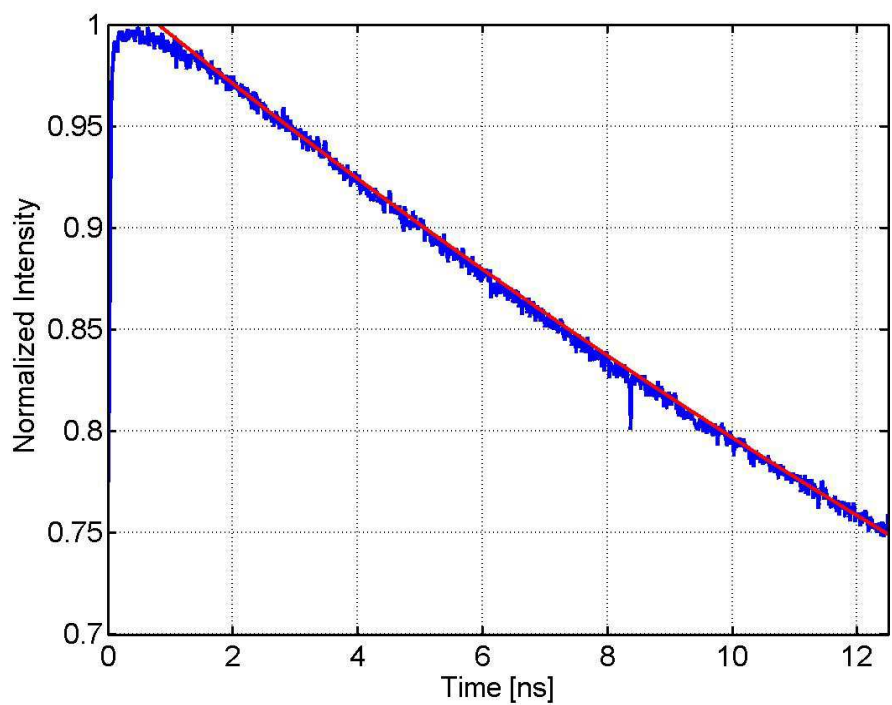

**Figure S4.** Luminescence decay trace (blue line) for quantum dots in water solution. The red line is a single exponential fit with a decay time of 40.5ns. No fast transients could be detected in this case.

**Antenna fabrication and QD binding area definition**

The Au antennas were defined by electron beam lithography in a FEI InspectF50 system at 30 keV acceleration voltage. Glass substrates were capped with a 10 nm conducting ITO layer deposited by electron beam evaporation. For the first lithography step the substrates were coated with a 120 nm PMMA layer and baked at 175°C for 5 min. During the first e-beam exposure, disk and disk dimer antennas were defined together with alignment markers for the second e-beam step. The sample was developed in a MIBK:IPA (1:3) mixture for 45 sec followed by immersion in IPA solution to stop the develop. The substrate was finally dried in a flow of nitrogen ($N_2$). A 40 nm Au film was thermally evaporated at a rate of 2 Å/s. The lift-off was performed in acetone at 55°C during 1 h followed by rinsing in IPA before drying with $N_2$. For the second e-beam step, the substrate was coated with a 50 nm PMMA layer and baked at 140°C for 5 min. The QD binding areas were defined by single pixel dot exposures. The design pattern was aligned to the previously defined alignment marks by means of a standard alignment layer of the ELPHY Plus software from Raith. The sample was developed in a MIBK:IPA (1:3) mixture for 45 sec followed by immersion in IPA. The substrate was finally dried with $N_2$ and ready for the monolayer formation and QD binding protocol. The holes dimensions were previously calibrated by fabricating a test sample with different single pixel exposures to which, after development, a 10 nm Ti layer was deposited inside the holes and subsequently lifted off in acetone. Figure S5 shows the positioning of 50 nm Ti dots aligned to the center of dimer rod antennas.

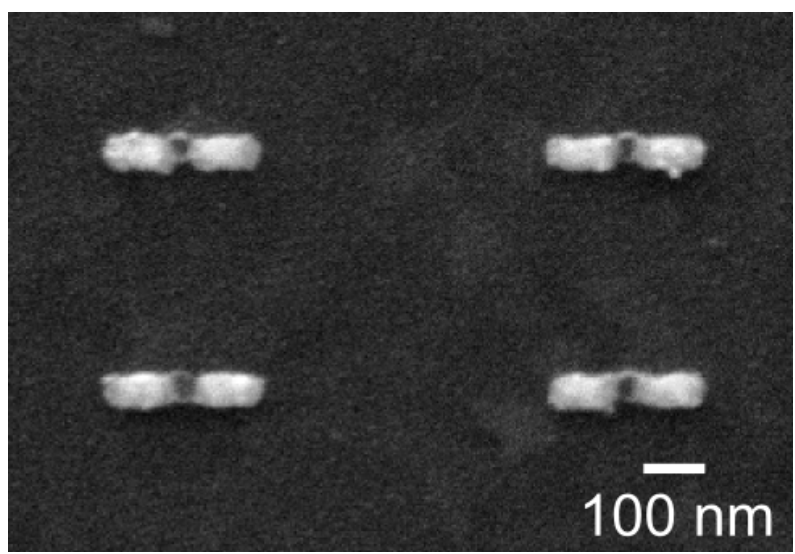

**Figure S5.** SEM image showing the positioning of Ti dots with a diameter of 50 nm placed at the center of the gap of a rod dimer antenna with a gap of 30 nm.

As one can clearly see from Figure S5, the area of Au uncovered on each rod by the second lithographic step is quite small in comparison to the total area exposed. Thus the judicial positioning of a single QD is a combination of a precisely controlled second lithography step and a controlled sequence of binding steps. Furthermore, the minimal exposed gold surface area is filled with a thiol monolayer (MUA), constituting many small molecules, which later is modified to capture a protein, namely bovine serum albumin (BSA) that has a reported hydrodynamic diameter of 8nm. Realistically, one to a few BSA molecules are only needed to completely coat this partial MUA monolayer and thus reduces further the possibility of multiple QD attachment. Finally, the probability and possibility of multiple QD attachments is further restricted due to steric hindrance. One must consider a 50nm diameter PMMA hole which is either partially or wholly filled with BSA protein. The activated QD in bulk solution must encounter the hole, penetrate therein and then encounter an amino group present on the BSA in order to create a covalent and strong bond between BSA and QD. Once formed, this initial QD hinders the approach of other QDs considering also the fixed hole diameter. This does not eliminate completely the possibility of multiple QDs, but largely restricts this event from happening.

**Photoluminescence decay traces**

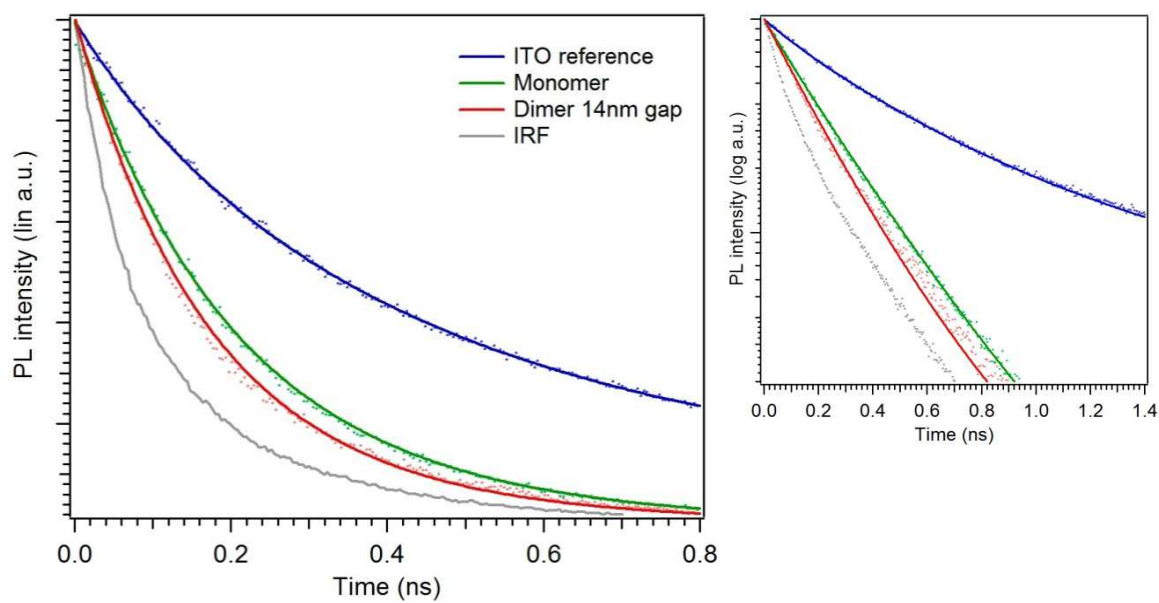

**Figure S6.** Comparison of luminescence decay traces taken at 20 µW average excitation power. The solid lines are numerical fits with a bi-exponential model convoluted by the IRF. Note the linear scale in the vertical axis, the inset displays the same traces on logarithmic scale.

**QD influence on the antenna's response**

In the absence of the QD, the excitation intensity enhancement for a 90 nm gold dimer antenna with 14 nm gap is computed to 250x (Figure S7a). However, when a single QD is introduced in the gap, the simulated excitation enhancement factor drops down to ≈ 50x. This strong influence is related to the high refractive index (n ≃ 2.47) and the large size (10 nm) of the QD.

We also investigate experimentally the position of the antenna extinction resonance wavelength measured on dimer antenna arrays with 2 µm pitch as the number of exposed antennas to the second lithography was increased. A 17 nm red-shift of the resonance wavelength is observed as the coverage density of QDs on the antenna array is increased from 0 (no QD) to ∼ 100% (all antennas bear at least one QD in their gap), see Fig. S7b.

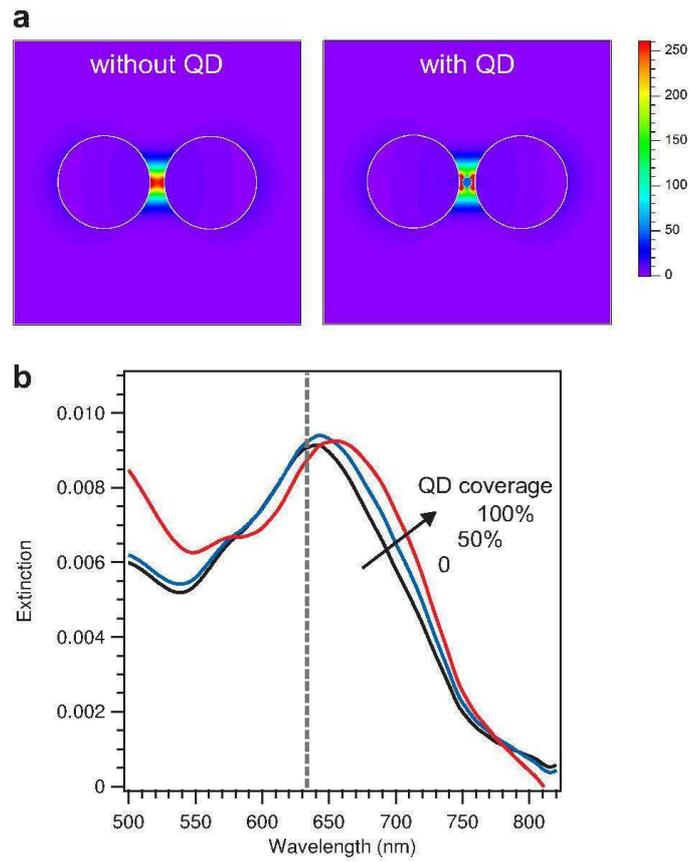

**Figure S7.** (a) Computed field intensity distribution on 90 nm gold dimer antenna with 14 nm gap, respectively without and with a 10 nm QD in the gap for λ = 636 nm. (b) Extinction spectra on dimer antenna arrays with 2 µm pitch for increasing QD coverage densities.